\documentstyle[11pt,newpasp,twoside,epsf]{article}
\def\edcomment#1{\iffalse\marginpar{\raggedright\sl#1\/}\else\relax\fi}
\reversemarginpar

\begin{document}
\title{The Formation of Galaxies, the Formation of Old Globular Clusters and the Link with High-Redshift Objects}
\author{Denis Burgarella, denis.burgarella@astrsp-mrs.fr}
\affil{Observatoire Astronomique Marseille-Provence, traverse du siphon, 13376 Marseille 
cedex 12, France}
\author{Markus Kissler-Patig, mkissler@eso.org}
\affil{ESO, Karl-Schwarzschild-Str. 2, 85748 Garching bei München, Germany}
\author{V\'eronique Buat, veronique.buat@astrsp-mrs.fr}
\affil{Observatoire Astronomique Marseille-Provence, traverse du siphon, 13376 Marseille 
cedex 12, France}

\begin{abstract}
 In this paper, we are exploring the properties of old, metal-poor globular 
 clusters in galaxies.  We investigate whether their properties are related 
 to the properties of their host galaxies, and whether we can constrain their 
 formation. The main result is that 
 the mean metallicities of old GC systems are found to lie in 
 a narrow range -1.7 $<$ [Fe/H] $<$ -1.1 (80 \% of the population). Moreover, no 
 correlations are found between the mean metallicities and other galaxy 
 properties which implies a GC formation independent of 
 the host galaxies. Further, we try to identify the sites of old, metal-poor 
 GC formation, with any currently known high redshift objects. 
 We find that the metalicities of damped Ly$\alpha$ systems in the redshift 
 range 1.6 $<$ z $<$ 4 
 are consistent with our GC metalicities, which suggests that 
 these high-density 
 neutral gas objects may be the progenitors of the old, metal-poor globular 
 clusters.
\end{abstract}


\section{Introduction}

Statistics is always a key-point in scientific studies and it is no surprise
that, as any scientist, the astronomer is looking for large, statistically
significant samples to properly analyze the Universe and its content.
Unfortunately, the intrinsic size of the Universe is turning this simple point
into a difficult brain teaser due to the faintness of the above objects.
Globular clusters are likely to contain some of the oldest known stellar
populations of the Universe.  As such, they potentially hold a cosmologically
significant information on their formation and more generally on the conditions
that prevailed more than 10~Gyrs ago.

In
brief, we have started a study that is heading at selecting the oldest globular
clusters (GCs) from the largest available sample of extragalactic GC systems.
Previous studies often assumed the systems as an homogeneous population and 
used the mean properties (metallicity) of the GC systems as the main
parameter.  Only in the most recent works GCs have been split up
in sub-populations (Forbes et al.  1997; C\^ot\'e et al.  1998). 
Going back to the very formation of the galaxies
(and maybe before), asks to make sure that only the oldest (i.e.  reliable fossils) GCs are picked up.  Our choice is to select the
metal-poor GCs.  Indeed, if we can find old metal-rich globular
clusters (Ortolani et al.  1995; Puzia et al.  1999), only in very limited cases
could we have a late formation of metal-poor GCs. An important step has
been to discover that galaxies other than our own contain metal-poor
sub-populations that can be associated to a halo component (Puzia et al.  1999).

A more detailed report of this work will be published elsewhere (Burgarella, 
Kissler-Patig \& Buat, 2000).

\section{The compilation of Old Globular Cluster Populations}

Our goal is to select the oldest GCs around galaxies and to
compare their metallicities with the host galaxy properties, as well as to 
compare the systems with each other. The compilation
includes galaxies of all types, however spiral galaxies are under-represented
while bright elliptical galaxies dominate the sample. The detection of
several peaks in the metallicity distribution function is always a problem and
we use the mixture-modeling algorithm (KMM) developed by Ashman et al.  (1994)
to detect and quantify the bimodality and estimate the mean metallicity of the
metal-poor GC populations around the sampled galaxies. 
This compilation of 38 GCs
systems includes galaxies of all types and our sample includes galaxies over 10
magnitudes in absolute brightness (see Burgarella et al. 2000).

\section{Mean metallicity against galaxy luminosity}

Before a clear separation of metal-poor and metal-rich populations could be
performed in other galaxies than in the Milky Way, the mean metallicities of the
whole GC system was thought to correlate with the galaxy luminosity (van den
Bergh 1975; Brodie \& Huchra 1991).  Actually, this apparent correlation was
mainly due to the fact that the the brightest galaxies are ellipticals which
have, on average, a higher GC mean metallicity than spirals and dwarfs
(e.g.~Ashman \& Zepf 1998; Gebhardt \& Kissler-Patig 1999).

The sample of GC systems presented in this paper is the largest
database to-date, and about 3 times more numerous than Forbes et al.'s (1997)
initial dataset.  The mean [Fe/H] lies at [Fe/H]$=-1.40\pm0.06$ with a
dispersion of $\sigma=0.24\pm0.05$, that is slightly more metal-poor on average
than, and exhibiting a scatter similar to the Forbes et al.  sample.
Fig.~1 shows the relative percentage of GC systems
within each bin of the metallicity function.  Indeed, the immediately apparent
result is that the mean-metallicities of metal-poor GCs are not
distributed at random:  most of them are lying around [Fe/H]$\sim-1.4$, with
64~\% within -1.5~$<$~[Fe/H]~$<$~-1.3 and 80~\% within
-1.7~$<$~[Fe/H]~$<$~-1.1.  We plot in Fig.~1 the metallicities of the
GC systems as a function of the absolute magnitude M$_V$.  {\it
The average metallicity of the metal-poor GCs is constant over a
very large range in absolute magnitude of the host galaxy ($-23<{\rm
M}_V<-16$)}.

\begin{figure} \plottwo{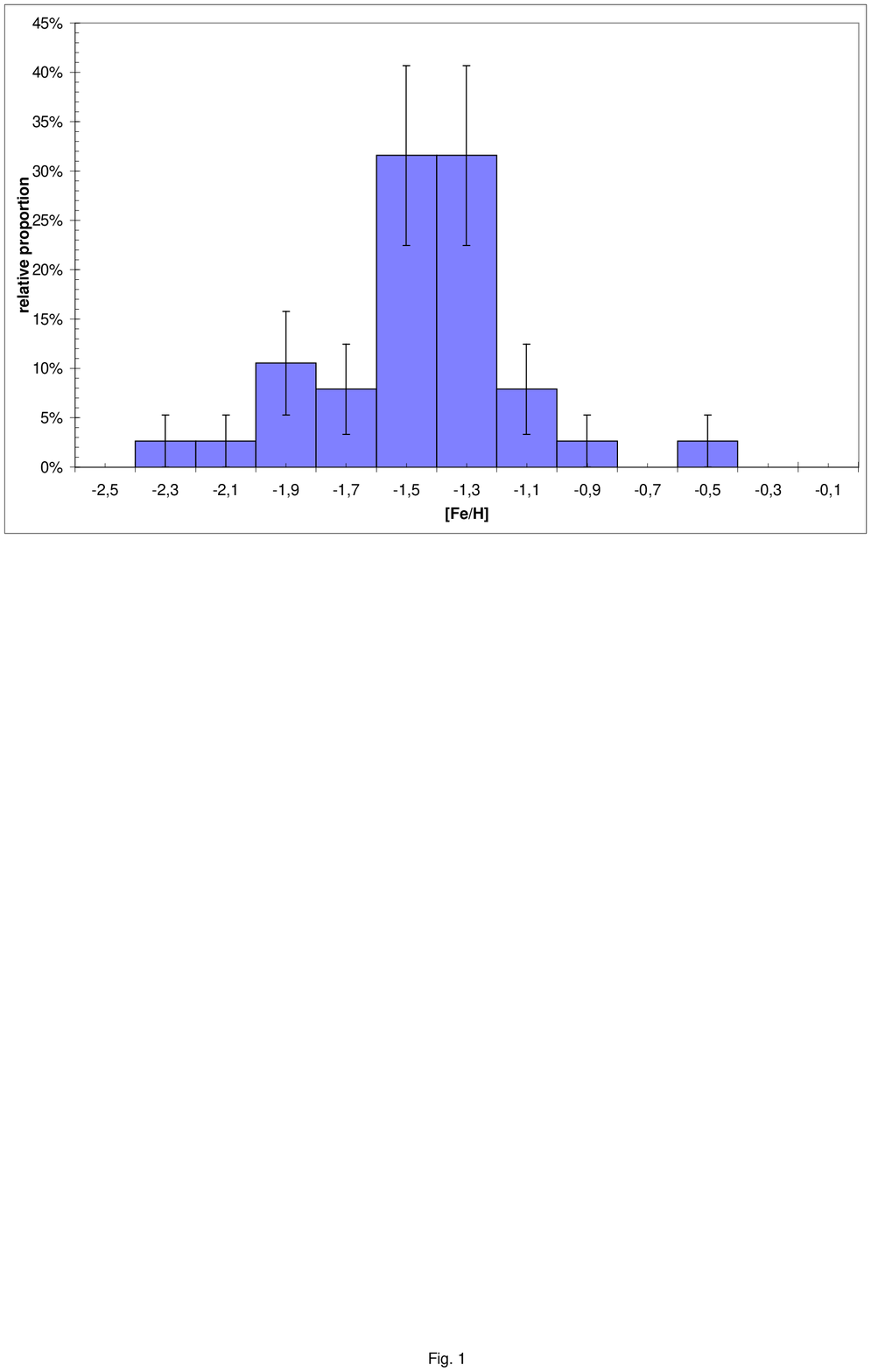}{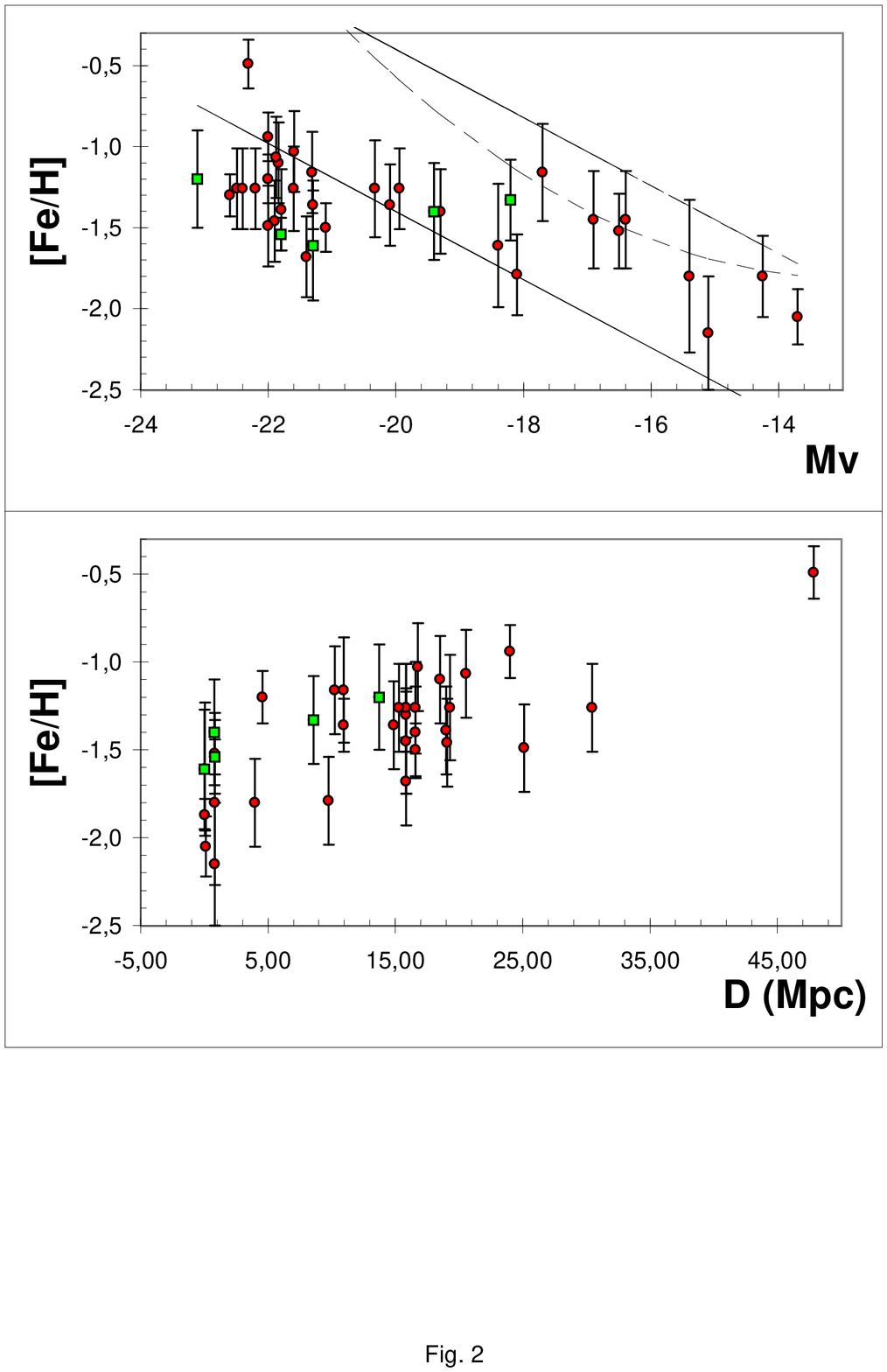} \caption{ 
a) Distribution of mean
metallicities for the GC system sample.  Note the narrow 
peak at
[Fe/H] $\approx$ -1.4 with 80~\% of the population within 
-1.7~$<$~[Fe/H]~$<$~-1.1.
b) Mean metallicity of the old,
metal-poor GC systems plotted against the absolute magnitude of
the parent galaxy M$_V$ and the distance to the MW (right).}
\end{figure}

\begin{figure} \plottwo{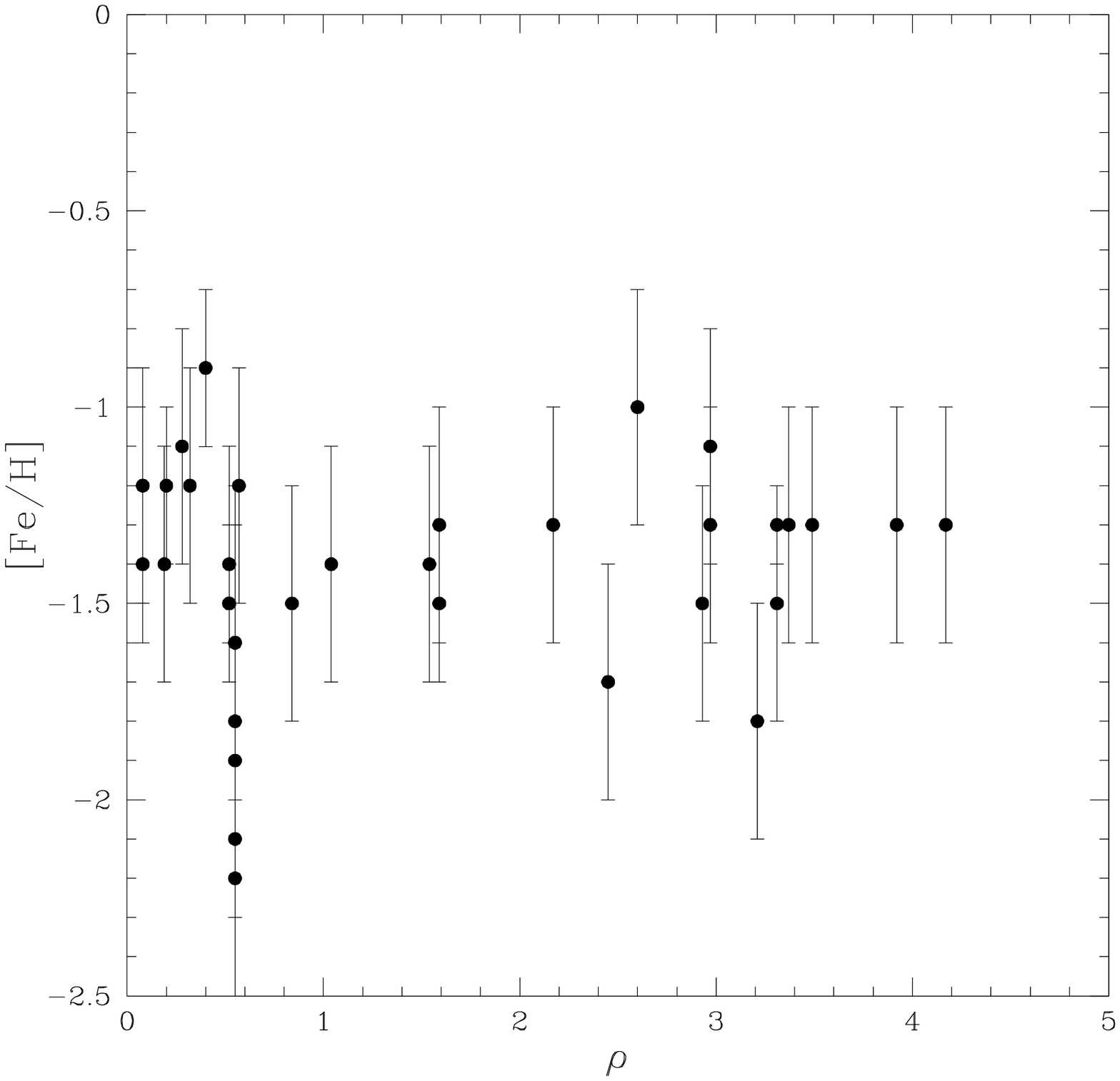}{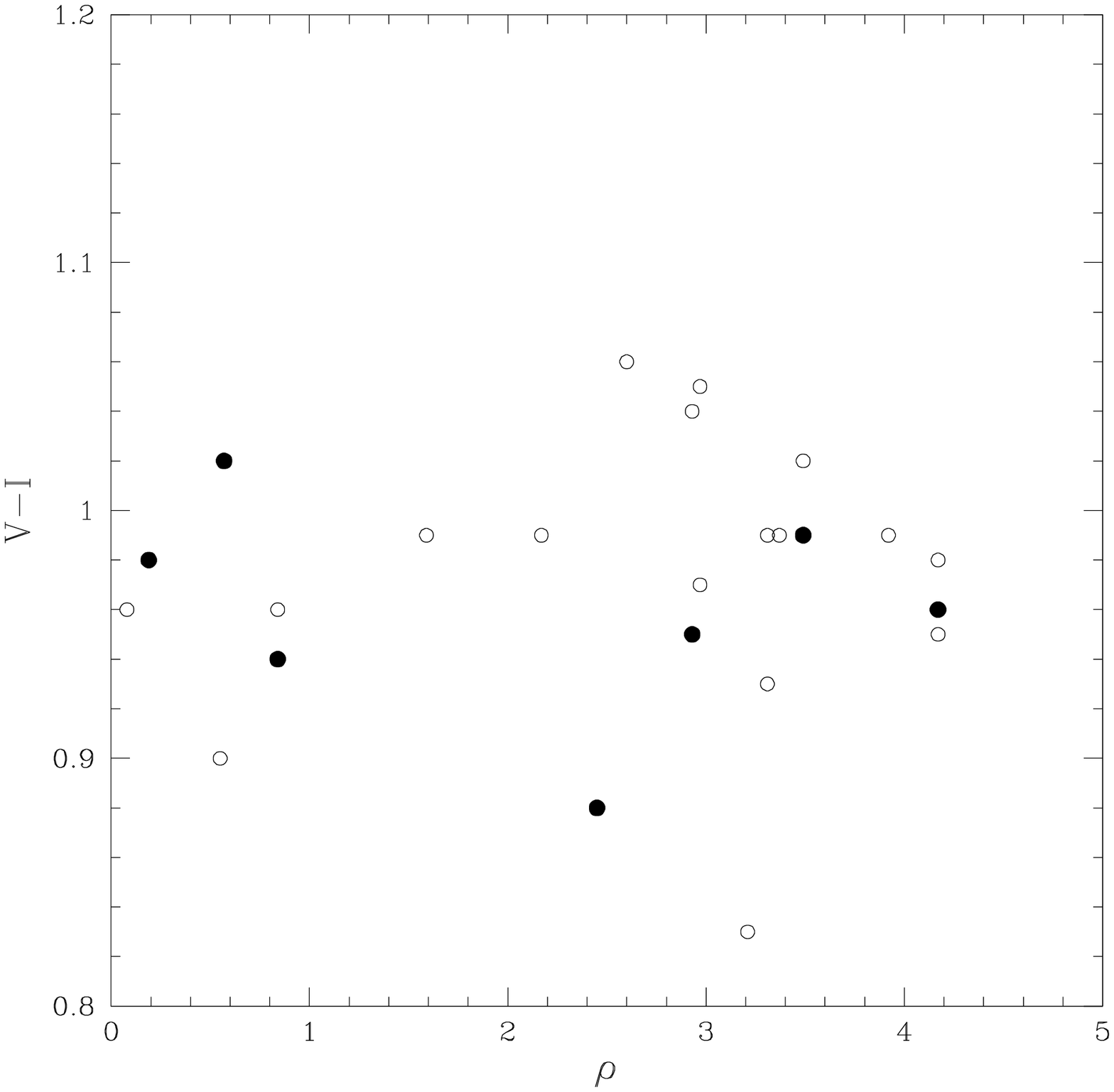} \caption{a) Mean 
metallicities and
b) peak $V-I$ colors of the metal-poor populations plotted against the host galaxy
environment density $\rho$.  Black dots are new values.}
\end{figure}

\section{Consequences on the formation scenarios of GCs, globular
cluster systems, and galactic halos}

Following Fall \& Rees (1988), GC formation models 
``{\it can be classified as primary, secondary or tertiary depending on whether GCs are assumed to form
before, during or after the collapse of proto-galaxies.}''.
It seems, however, that the borderline between the three classes is not always
very clear.  To better identify the origins of GCs, we prefer to
split the GCs on whether they are {\it external} to the
galaxy, and not associated with the {\it final} host galaxy, or whether they
formed {\it internally}, i.e.  are associated in some form with the {\it final}
host galaxy.  This terminology is relatively unambiguous if we specify that
pre-galactic fragments are not considered to be galaxies.  And since we consider
only old, metal-poor GCs assumed to have formed before or early in
the galaxy formation process, we do not take into account mergers of already
formed galaxies.

Now, if we concentrate our attention on the Milky Way GC system,
it seems that halo GCs of the Milky Way host (at least) two
populations that Zinn (1993) distinguished from their horizontal branch
types.  He called them 'old' and 'younger' halo GCs.  There are
hints for a similar differentiation in M33 (Ashman \& Bird 1993).  A probable
internal halo old GC population which would have formed internally
in the early galaxy lifetime by a dissipative collapse in a few Myr, and an
external halo population which would have formed around other satellite galaxies
and accreted afterwards.  If such a complex GC formation history
is valid for our own Galaxy and its nearest neighbors, it cannot be ruled out
for other galaxies either.

We could retain that~:  i) the mean metallicity of
halo GCs is independent of the host galaxy properties (M$_V$,
type, environment (Fig.~2), metallicity) and ii) halo GC populations have
very similar mean metallicities in all galaxies.  These two points can be added
to the dynamical information available for a number of metal-poor outer globular
clusters that tends to show that these clusters are on tangentially biased
orbits, as opposed to radially biased orbits expected if they had formed in a
collapse (Eggen et al.  1962). 

The bottom line from the above facts, is that the early cluster and star
formation was remarkably homogeneous in the local universe (within several tens
of Mpc).  The first collapsing fragments were extremely similar in mass and
abundances over large scales and collapse in very similar fashions independently
of the potential well (dark halo) in which they were located.  Presumably, the
distinction between galaxy types only appeared after the first formation of
stars and clusters in fragments.

\section{Time and sites of formation of the metal-poor GCs}

\subsection{The measurement of metallicity at high redshift} 

In this section, we will look for measurements of metallicity at high redshift
in order to compare with the average metallicity of our old GCs.
Indeed, if the GC formation is the first stellar formation episode
of what will become a new galaxy, the first-formed stars might have kept a
memory of the genuine intergalactic medium as it was before the galaxy
formation.  Among the objects observed at high redshifts for which the
metallicity can be estimated, two of them seem of interest:  Damped Ly$\alpha$
systems (DLAs), and Lyman Break Galaxies (LBGs).  Both the typical column
density in H{\tt I} and the observed metallicities for DLAs and LBGs are plotted
in Fig.~3, together with the same quantities for GCs (and the
Lyman $\alpha$ forest for completeness).  DLAs (Pettini et al.  1997) give a
measurement for the metallicity as a function of redshift of high density
neutral gas objects.  Lyman Break Galaxies (Steidel et al.  1996) can, in
addition, be used to estimate the star formation rate as a function of redshift.
Assuming a metal ejection rate (Pettini et al.  1997),
we can infer a chemical evolution of the Universe and compare it with other
estimates, in particular with the mean metallicity of the metal-poor globular
cluster systems.

\begin{figure} \plottwo{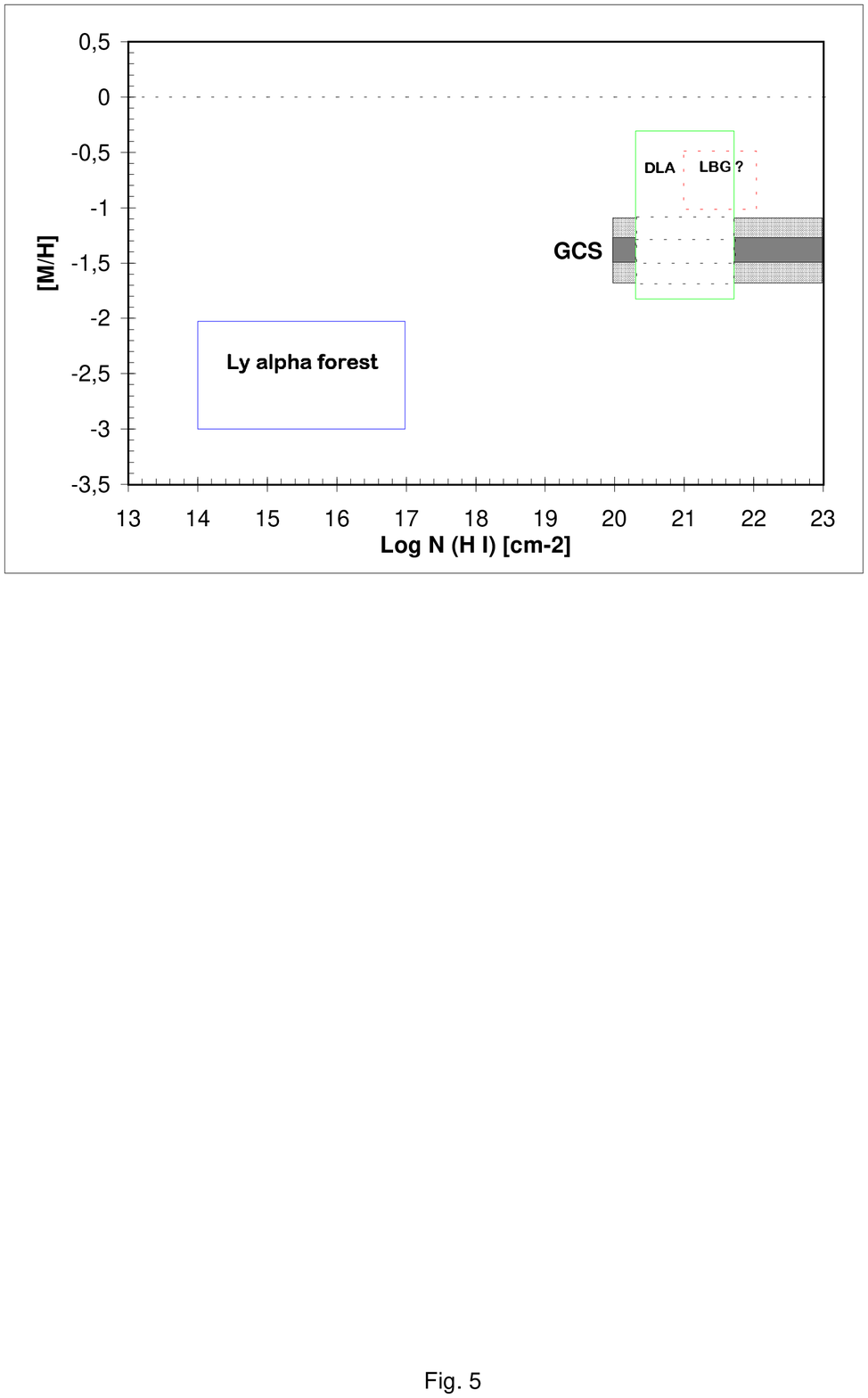}{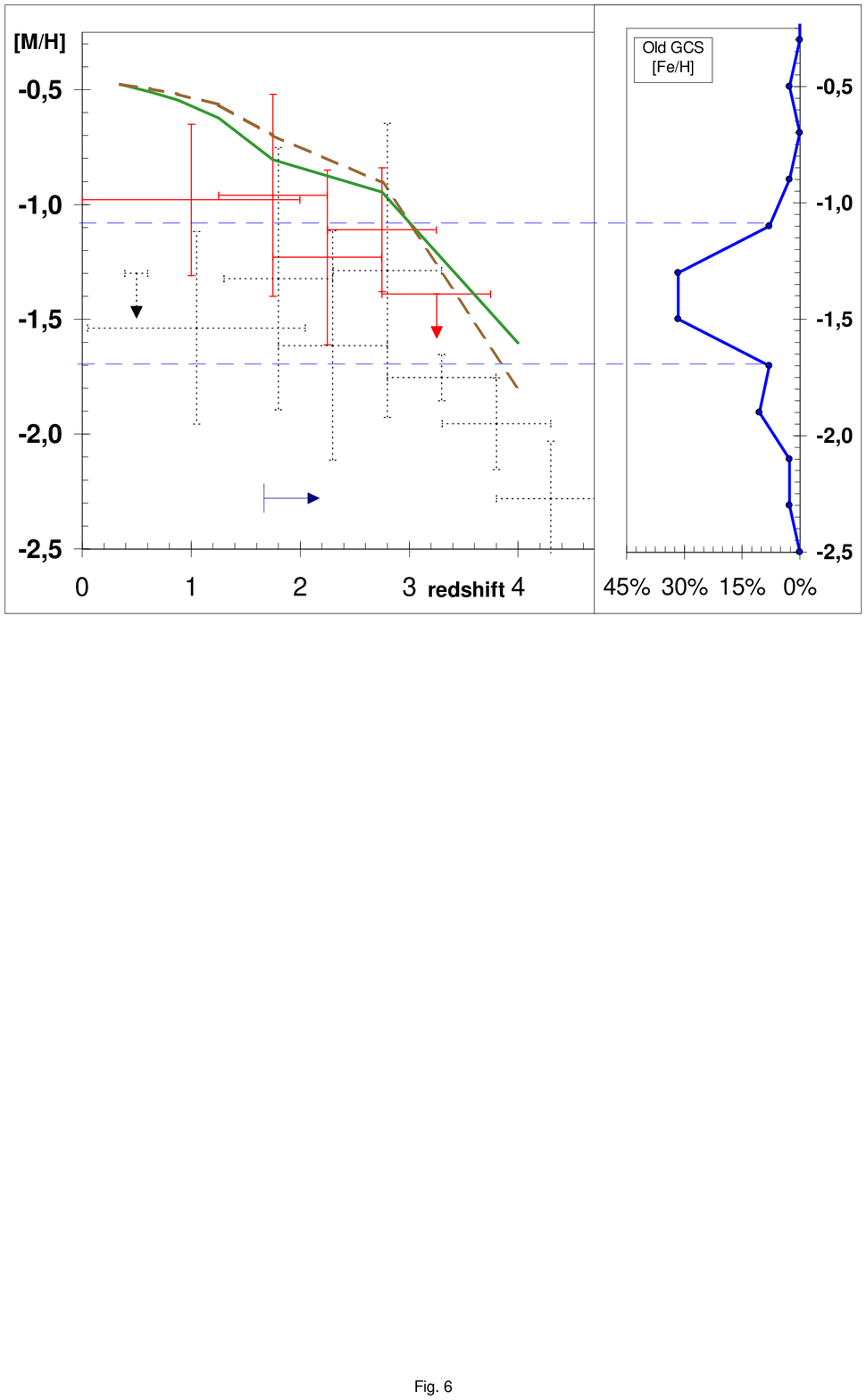} 
\caption{
a) This figure gives the rough location in 
metallicity against column density of neutral hydrogen for different components 
of the high redshift universe. We have added the results from our GC  
systems (80 \% and 64 \%) taking into account a low threshold 
N(HI) $\approx$ 20 for the star formation to occur (left side).
b) Comparative variation of the
metallicities with the redshift (H$_o$ = 50 km.s$^{-1}$.Mpc$^{-1}$ and q$_o$ =
0.5).  Right hand panel:  GC system metallicity distribution.
Left hand panel~:  the limits at 80 $\%$ of the GC system
mean-metallicity distribution are reported as dashed
lines.  The uppermost curves have been deduced from Steidel et al.'s (1999, 
Fig.9) star formation history (continuous line); the dashed line includes Barger et
al.  1999 FIR data.  [Zn/H] (continuous line crosses) and [Fe/H] (dashed
crosses) values of DLAs are taken from Burgarella et al (2000).  The age of 
the oldest Galactic
GCs are reported as an horizontal right-bound arrow.  Pettini et al.
1997 noted that assuming q$_0$ = 0.01 would shift the [M/H] by a factor of 2.
}
\end{figure}

However, [Fe/H] may not be a reliable estimate of the
metallicity of DLAs, since some Fe may be locked up in dust and thus the
measured [Fe/H] too low.  Pettini et al.  (1997) showed that [Zn/H] is a more
reliable estimator because it essentially measures the metallicity independently
of dust depletion.  From a [Zn/H] analysis of 34 DLAs, Pettini et al. (1997)
showed that z $>$ 1 DLAs are generally metal-poor (log ($<Z>/Z_\odot$) $<$ -1.0)
with a possible trend for z $>$ 3 DLAs towards a lower metallicity.  However,
the value at z = 3 is an upper limit and we would need better high redshift
values.  Although the dust depletion problem may make the direct use of DLA
[Fe/H] questionable, we try here to use its variation with redshift in order to
compare it with the information from GC systems.  The compilation is given in 
Burgarella et al. (2000). The [Zn/H] and [Fe/H]
variations as a function of the redshift are plotted in Fig.~3.  We use the
column-density weighted abundances~:  $ \rm [<M/H_{DLA}>] = log <(M/H)_{DLA}> -
~log (M/H)_\odot $ where $\rm <(M/H)_{DLA}$ (M = Fe or M = Zn) and the
associated standard deviations as defined in Pettini et al.  (1997).

The data presented in Fig.~3 can be used to constrain the GC
system formation.  In the first place, the analysis of the CMDs of old Galactic
GCs suggests the age of halo GCs to be more than 10
Gyr which corresponds to a GC formation not later than z $\sim$
1.6 (H$_0=50$ km.s$^{-1}$.Mpc$^{-1}$ and q$_0=0.5$).  On the other hand, the 
chemical evolution of
DLAs and LBGs is below the lower limit for 80 \% of our metal-poor globular
clusters at z $\approx$ 4.  The conclusion suggested by these data is that the
GC formation occurred in average in the redshift range 1.6 $<$ z
$<$ 4 (i.e approximately in the range 10 $<$ age (Gyrs) $<$ 12 with the assumed
cosmology).

From the above discussion we retain that DLAs (and LBGs) have approximately the
same range of metallicities and are observed in the redshift range expected for
the formation of metal-poor GCs.  Note, however, that DLAs contain
neutral gas while LBGs are star-forming objects.  As already suggested by Fynbo
et al.  (1999), we may wonder whether we are not observing the same objects at
different location in space or in time.  For instance, DLAs would be the source
of dense gas out of which old GCs formed while LBGs would be
star-forming regions e.g.  spheroids as proposed by Giavalisco et al.  (1996)
and Steidel et al.  (1996) but also surrounding fragments in the same potential
well which are only directly visible at high redshift when the star formation
turns on. Eventually these
fragments would be accreted by the large galaxy to produce a MW-like object.

\acknowledgements{We would like to thank K. Gebhardt for his help in handling
the data of the metal poor GCs and M. Pettini for helful discussions.}

\end{document}